\documentclass[draft]{agujournal2019}
\usepackage{url} 
\usepackage{lineno}
\usepackage[inline]{trackchanges} 
\usepackage{soul}
\draftfalse

\journalname{XXX}

\begin{document}

%
%


\title{Cracking the Code of Arctic Sea Ice: Why Models Fail to Predict Its Retreat?}

%
%




\authors{Ruijian Gou\affil{1,2}, Gerrit Lohmann\affil{2,3}, Deliang Chen\affil{4}, Shiming Xu\affil{4}, Ruiqi Shu\affil{4}, Shaoqing Zhang\affil{1,5}, Lixin Wu\affil{1,5}}

\affiliation{1}{Laoshan Laboratory, Qingdao, China.}
\affiliation{2}{Alfred Wegener Institute, Bremerhaven, Germany.}
\affiliation{3}{University of Bremen, Bremen, Germany.}
\affiliation{4}{Tsinghua University, Beijing, China.}
\affiliation{5}{Ocean University of China, Qingdao, China.}





\correspondingauthor{Ruijian Gou}{rgou@foxmail.com}




%
%

%
%


\begin{abstract}
Arctic sea ice is rapidly retreating due to global warming, and emerging evidence suggests that the rate of decline may have been underestimated. A key factor contributing to this underestimation is the coarse resolution of current climate models, which fail to accurately represent eddy–floe interactions, climate extremes, and other critical small-scale processes. Here, we elucidate the roles of these dynamics in accelerating sea ice melt and emphasize the need for higher-resolution models to improve projections of Arctic sea ice.
\end{abstract}

%
%

%


%
%
%
%

Arctic sea ice is one of the most critical components of the climate system. Due to global warming and Arctic amplification—where warming in the Arctic occurs at roughly four times the global average \cite{smith2019polar}—Arctic sea ice is retreating rapidly and may disappear entirely during summer by the mid-21st century \cite{massom2018antarctic}. However, many of these projections rely on low-resolution climate models that do not resolve essential processes such as eddy–floe interactions, meltwater dynamics, and fine-scale thermodynamics. For instance, the CMIP model ensemble underestimates the observed summer sea-ice loss in recent years \cite{shu2020assessment, stroeve2007arctic}, while higher-resolution models tend to simulate smaller sea ice extents \cite{docquier2019impact, selivanova2023past, chang2020unprecedented}. 

The discrepancy between low- and high-resolution model outputs remains poorly understood, partly due to observational uncertainties and the intrinsically multiscale nature of sea-ice formation and melt processes \cite{notz2020arctic, jahn2024projections} , which range from millimeters to thousands of kilometers. Each spatial scale can play an independent and unique role \cite{golden2020modeling}, presenting major challenges for simulating sea ice in coarse-resolution models. While improving the representation of sea-ice physics remains important, increasing model resolution—now increasingly feasible due to growing computing power—should also be a priority. Finer resolution not only improves the fidelity of simulated physics but also allows explicit representation of critical small-scale processes. This perspective highlights how higher-resolution models can enhance the representation of key physical processes governing Arctic sea ice loss and implies that current projections may underestimate the rate of future retreat.

\section{Unresolved ocean eddies and floes}
Even the most advanced numerical models used in climate projections have grid sizes too coarse to resolve key processes in the marginal ice zone—areas partly covered by sea ice—where ocean eddies and sea-ice floes play a dominant role. Floes, which vary widely in size, often fall below model grid scales. Even the highest-resolution models treat sea ice as a continuum, unable to resolve even the largest \cite{gupta2024eddy}. Simply increasing resolution does not guarantee better representation of floe dynamics, especially when floe scales approach the model’s grid spacing—a limitation rooted in current numerical schemes.

Arctic ocean eddies typically span a few kilometer \cite{liu2024spatial}, yet most climate models use grid resolutions ranging from tens to hundreds of kilometers. This coarse resolution necessitates parameterizations for eddy-induced transport \cite{smith2019polar}. While high-resolution models can resolve mesoscale eddies at lower latitudes and deactivate parameterizations accordingly, Arctic eddies in marginal seas remain largely unresolved. Even at the highest feasible resolutions, Arctic eddy activity must be inferred rather than simulated directly.

The interaction between sea-ice floes and ocean eddies is essential for accurately projecting long-term sea-ice changes. At the ice edge, meltwater enhances lateral density gradients that drive eddy formation, while sea ice dissipates eddies \cite{horvat2016interaction}. As sea ice retreats, eddy activity is expected to increase, with generation outpacing dissipation \cite{li2024eddy}. Smaller floes reduce dissipation \cite{gupta2024eddy, horvat2016interaction}, and thus, higher-resolution models that resolve smaller floes may project more vigorous eddy activity. These eddies can transport heat into ice-covered regions, where smaller floes are especially susceptible to lateral heat fluxes \cite{gupta2024eddy, horvat2016interaction}. Together, these processes imply that sea ice retreat may be faster in high-resolution models than in coarse-resolution projections.

\section{Underestimated climate extremes}
Higher-resolution models provide more degrees of freedom, enabling greater variability across spatial and temporal scales \cite{laepple2023regional}. As a result, they better capture the tails of probability distributions and simulate stronger, more realistic climate extremes \cite{carleton2022valuing, contzen2023long}.

Among these extremes, atmospheric rivers—narrow bands of intense moisture transport—have emerged as significant contributors to sea-ice melt. Their frequency and intensity have increased in recent decades, enhancing downward longwave radiation and accelerating surface melt \cite{zhang2023more}. Projections indicate that extreme atmospheric rivers will become more common in a warming climate. Higher-resolution models simulate a stronger increase in their intensity \cite{wang2023extreme}, implying that coarse-resolution models likely underestimate sea-ice loss from these events.

Similarly, intense polar cyclones play a major role in sea-ice dynamics \cite{massom2018antarctic}. These storms can generate long-period swells that fracture large floes and expose more open water, promoting both lateral and basal melt \cite{zhu202312}. These processes are better resolved in high-resolution models, which again point to more pronounced future sea-ice retreat.

Other climate extremes, while less well understood, may also affect sea-ice loss. Paleoclimate evidence shows that millennial-scale temperature fluctuations can cause substantial ice sheet retreat over 10,000-year periods due to nonlinear mass balance responses \cite{niu2019climate}. It is unclear whether sea ice responds symmetrically to extreme fluctuations. Under background warming, cold extremes are unlikely to promote sea-ice growth. Instead, Arctic cyclones increasingly skew the distribution toward warm extremes near the surface \cite{parker2022influence}, and marine heatwaves are becoming more common in the marginal ice zone \cite{hu2020marine}. Sea ice is likely to respond most strongly to warm extremes, further accelerating its decline.

\section{Other fine-scale processes}
Even with increased resolution, some small-scale processes—such as surface and internal waves—remain unresolved and must be parameterized. Despite their scale, these processes significantly influence sea-ice thermodynamics. Ocean surface waves can break and perturbsea ice both dynamically and thermodynamically \cite{zhu202312, casas2024wind}. Their impact intensifies when floes are better resolved, due to reduced energy dissipation and greater lateral heat flux sensitivity \cite{gupta2024eddy}. Internal wave-driven mixing may deliver additional heat to the ice base. This process is often treated as a constant in models, yet internal wave activity is expected to increase as sea ice melts \cite{hartharn2024interactions}. Capturing this feedback will be key for accurate future projections.

Moreover, coarse-resolution models struggle with larger-scale but spatially localized processes. For example, they substantially underestimate oceanic heat transport through the Bering Strait—a narrow but critical channel connecting the Pacific and Arctic Oceans \cite{xu2024high}. This underestimation arises from poor resolution of boundary currents and narrow passages, leading to an underprediction of regional sea-ice loss. 

\begin{figure}
\noindent\includegraphics[width=\textwidth]{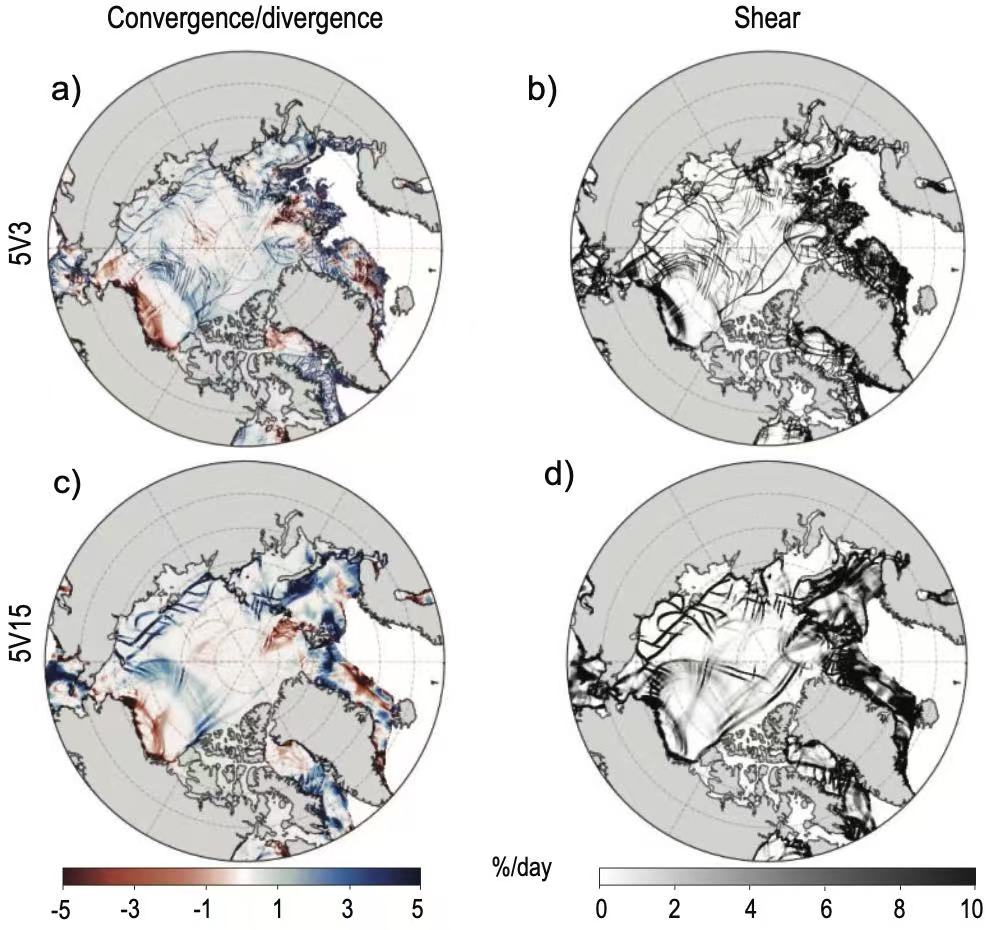}
\caption{Distributions of sea ice convergence (a, c) and shear rates (b, d) from model simulations with horizontal resolutions of 1.4 km (a, b) and 7 km (c, d). Reproduced from \citeA{zhang2023toward}, CC BY 4.0.}
\label{figure_1}
\end{figure}

Sea-ice leads and polynyas—linear fractures and open-water areas that induce intensive air-sea interaction and sea ice mass balance changes—are also poorly represented in coarse-resolution models. These features result from plastic failure and stress-induced fracturing, forming networks of cracks, ridges, and openings. High-resolution models can simulate these patterns with increasing realism. Figure \ref{figure_1} compares 7 km and 1.4 km model simulations. While both resolve large-scale drift patterns, the 3 km simulation captures a much finer network of linear kinematic features, particularly in the western Arctic Basin and Beaufort Sea. Shearing zones, leads, and deformation structures appear more coherent and dynamically consistent. Notably, the Beaufort Gyre circulation and related strain zones are more realistically represented.

These kilometer-scale features are not just visually detailed—they are physically crucial. They influence momentum transfer, air–sea heat exchange, and brine rejection, all of which feed back into the broader ocean–atmosphere system. Accurately simulating these dynamics is essential for both seasonal forecasting and long-term climate projections. Continued progress in high-resolution modeling is vital to understanding how such fine-scale processes shape Arctic and global climate responses \cite{zhang2023more}.

\section{Sea ice rheology and future high-resolution sea ice modelling}
Sea ice behaves as a non-Newtonian fluid and is distinct from both the atmosphere and ocean. The Viscous-Plastic (VP) model introduced by \citeA{hibler1979dynamic} was the first rheological framework used in large-scale sea ice simulations. This dynamic–thermodynamic model, combining continuum-based sea-ice dynamics with thermodynamics to account for heat exchange, became foundational in oceanography and remains a standard approach in sea ice modeling. 

Since then, various rheology models have been proposed and implemented in large-scale simulations, including anisotropic rheology \cite{tsamados2013impact} and brittle rheology frameworks such as the Maxwell-Elasto-Brittle (MEB) model \cite{dansereau2016maxwell}. Many of these models successfully reproduce observed multifractal sea ice deformation patterns \cite{bouchat2022sea, xu2021comparison}. However, they all share a core limitation: treating sea ice as a continuous medium. Consequently, their spatial resolution is typically coarser than the size of individual sea ice floes and cannot resolve floe-scale processes. 

As grid resolution approaches the scale of individual floes, the assumptions underpinning Hibler’s continuum-based VP model begin to break down \cite{feltham2008sea}. At this finer scale, discrete interactions between floes—such as collisions, rafting, and fragmentation—cannot be captured accurately. This limitation makes the VP model insufficient for simulating phenomena in the marginal ice zone, where floe-scale dynamics dominat \cite{hopkins1991numerical}. 

Emerging approaches like the Discrete Element Method (DEM) are better suited for high-resolution simulations that explicitly resolve individual floes. For instance, \citeA{damsgaard2018application} developed DEM-based models that simulate mechanical interactions between floes, significantly improving the representation of heterogeneous and fragmented ice fields. While the VP model remains valuable for large-scale simulations, increasing computational capacity now supports the development of models that resolve floe-scale behavior, enhancing simulation accuracy—especially in marginal ice zone.

Several alternative methods have been proposed to simulate floe-level processes. DEM treats sea ice as a collection of rigid bodies, enabling simulations of dynamic interactions such as collisions and rafting—particularly effective in marginal ice zones \cite{hopkins1996mesoscale}. Another method, Smoothed Particle Hydrodynamics (SPH), models sea ice as a set of interacting particles, allowing simulations of ice fragmentation and wave–ice interactions, including floe breakup \cite{herman2016discrete}. A third approach, Floe Size Distribution (FSD) models, represents floe evolution statistically, offering a computationally efficient method for simulating the bulk effects of floe dynamics without resolving individual floes \cite{roach2018consistent}. 

Each of these methods involves a trade-off between computational efficiency and physical detail. A logical next step is the development of particle-based methods that simulate sea ice as a system of interacting particles, capable of representing floe dynamics in both pack ice and marginal ice zones.

\section{Lattice Boltzmann methods and sea ice thermodynamics}
The Lattice Boltzmann Method (LBM) \cite{wolf2004lattice}, originally developed for simulating fluid dynamics, has recently been applied to large-scale ocean circulation \cite{lohmann2021mathematics}. This intersection of mathematical methods and climate modeling underscores the potential of alternative frameworks for resolving complex climate system behavior. \citeA{freitag1999untersuchungen} used LBM to model brine channel transport within the microstructure of sea ice. While applications of LBM to large-scale sea ice dynamics remain limited compared to traditional continuum models, the method shows promise for resolving fine-scale processes in complex geometries. 

Future advances in high-resolution sea ice modeling are expected in the area of thermodynamics, particularly processes like brine transport, mushy-layer growth, and convection \cite{notz2009desalination, vancoppenolle2010modeling}. These processes affect the phase behavior of sea ice and the development of brine channels \cite{wells2019mushy}, and they inform refined thermodynamic models applicable to large-scale systems \cite{vancoppenolle2010modeling, turner2013two}. To date, the parameterization of turbulent heat fluxes to the sea ice is still based on the bulk flux algorithm. As a practical first step for an improvement, a parametrization based on the theory of maximum entropy production (MEP) has been indicated to more accurate \cite{zhang2021modeling, wang2014model}. It is a statistical approach built upon probability theory and thus does not depend on other climate variables, which could significantly reduce uncertainty.

\section{Machine learning as a bridge across scales}
To better understand and simulate the multiscale complexity of Arctic sea ice, high-resolution models must be complemented by emerging computational strategies—especially machine learning (ML). In recent years, Earth system science has increasingly used ML to parameterize small-scale processes that are unresolved or poorly represented in traditional models \cite{bracco2025machine}. These ML-derived parameterizations, trained on satellite data, field observations, or high-resolution simulations, can be integrated into physical models to improve their accuracy while retaining interpretability.

Recent research has shown the potential of ML-based modules in modeling submesoscale eddies \cite{bolton2019applications, zanna2020data}, where traditional analytic closures fail. These modules can function as efficient and stable “plug-ins” in climate models, enhancing the simulation of local transport and mixing without significant computational cost. Explainable AI techniques further allow researchers to interpret the structure and behavior of these ML models \cite{chen2024collaboration}, making them tools not only for prediction but also for scientific insight.

\section{Toward a more realistic Arctic future}
The convergence of high-resolution modeling techniques—including DEM, SPH, FSD, and LBM—is unlocking new insights into Arctic sea ice dynamics. By explicitly representing processes that were previously overlooked or misrepresented, these models consistently suggest a faster rate of sea ice decline than current projections based on coarse models. This accelerated loss is not merely an Arctic phenomenon; it has cascading effects on global albedo, air–sea heat exchange, ocean buoyancy, and atmospheric circulation.

Capturing small-scale processes and integrating them into global climate frameworks is thus not just a technical challenge—it is a scientific imperative. As the Arctic trends toward a seasonally ice-free state, the fidelity of projections depends on representing the full spectrum of physical processes, from sub-meter-scale floe interactions to basin-wide currents. By increasing spatial resolution and embedding detailed physics—augmented by machine learning—researchers can move toward more accurate and actionable climate projections.

\begin{figure}
\noindent\includegraphics[width=\textwidth]{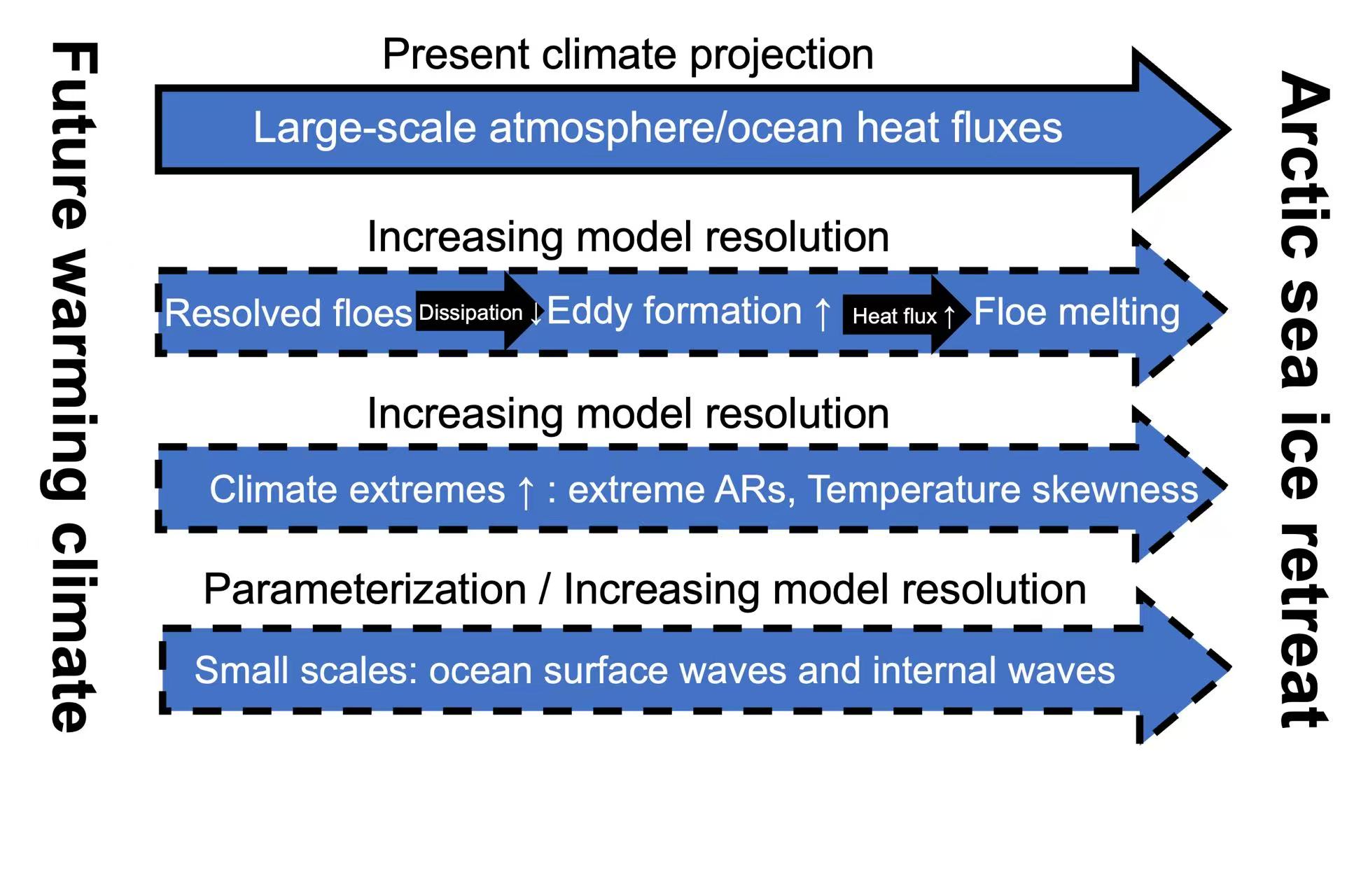}
\caption{Schematic illustrating the pathways through which future climate warming contributes to Arctic sea ice retreat. Dashed arrows represent additional pathways enabled by the development of higher-resolution climate models. AR and MHW refer to atmospheric river and marine heatwave, respectively.}
\label{figure_2}
\end{figure}

\newpage
\bibliography{agusample}

%
%
%
%
%

\end{document}